\documentstyle[12pt]{article}
\newcommand{\be}{\begin{equation}}
\newcommand{\ee}{\end{equation}}
\newcommand{\bea}{\begin{eqnarray}}
\newcommand{\eea}{\end{eqnarray}}
\newcommand{\ba}{\begin{array}}
\newcommand{\ea}{\end{array}}

\def\bbox{{\,\lower0.9pt\vbox{\hrule \hbox{\vrule height 0.2 cm
\hskip 0.2 cm \vrule height 0.2 cm}\hrule}\,}}
\newcommand{\dsl}{\pa \kern-0.5em /}

\newcommand{\nn}{\nonumber \\}

\font\mybb=msbm10 at 10pt
\def\bb#1{\hbox{\mybb#1}}

\def\bR {\bb{R}}
\def\bE {\bb{E}}

\def\bC {\bb{C}}

\begin{document}



\begin{titlepage}
\rightline{DAMTP-2004-134}
\rightline{\tt{hep-th/0411206}}

\vfill

\begin{center}
\baselineskip=16pt
{\Large\bf  Field Theory Supertubes{$^\star$}}
\vskip 0.3cm
{\large {\sl }}
\vskip 10.mm
{\bf ~Paul K. Townsend}
\vskip 1cm
{\small
Department of Applied Mathematics and Theoretical Physics,\\
Centre for Mathematical Sciences, University of Cambridge, \\
Wilberforce Road, Cambridge CB3 0WA, United Kingdom 
}

\end{center}
\vfill

\par
\begin{center}
{\bf ABSTRACT}
\end{center}
\begin{quote}

Starting with intersecting M2-branes in M-theory, the
IIA supertube can be found by  $S^1$ compactification 
followed by a boost to the speed of light in the 11th dimension. 
A similar procedure applied to Donaldson-Uhlenbeck-Yau instantons 
on $\bC^3$, viewed as intersecting membranes of $D=7$ supersymmetric Yang-Mills
(SYM)  theory, yields (for finite boost) a new set of 1/4 BPS equations for 
$D=6$ SYM-Higgs theory, and (for infinite boost) a generalization of the 
dyonic instanton equations of $D=5$ SYM-Higgs theory, solutions of which 
are interpreted as Yang-Mills supertubes and realized as configurations 
of IIB string theory.

\vfill
  \hrule width 5.cm
\vskip 2.mm

\end{quote}

$\star$ Contribution to ``Strings 2004''.
\end{titlepage}

\setcounter{equation}{0}
\section{Introduction}

Starting with the 1/4 supersymmetric intersection of two M2-branes in
M-theory, one can obtain other 1/4 supersymmetric configurations. 
For example, compactify on the 11th dimension to get the 
1/4 supersymmetric configuration in which a IIA string ends on 
a D2-brane. Now boost in the 11th dimension; in ten 
dimensions this corresponds 
to adding D0-charge so that the IIA string now ends on a 
bound state of a D2-brane with dissolved 
D0-branes. What happens if we boost to the speed of
light in the 11th dimension? Consider this question at the level 
of the effective Dirac-Born-Infeld (DBI) theory for the D2-brane, which is
just a dual version of the 11-dimensional supermembrane. At finite
boost, we have a `dyonic BIon', which is a D2-brane spike carrying 
constant electric flux and a constant magnetic charge 
density \cite{Gauntlett:2000de}. 
As we boost to the speed of light, the spike becomes more tubular and
we end up with a supertube \cite{Mateos:2001qs}. 

In effect, we have constructed the D2-brane supertube from the
11-dimensional supermembrane (the ``M-ribbon'' is an alternative
starting point \cite{Hyakutake:2002fk}), but 
the worldvolume action for the
supermembrane exists in spacetime dimension $D=4,5,7,11$ 
\cite{Bergshoeff:1987cm}, and the
above (worldvolume) construction works as well for $D=5,7$ as it 
does for $D=11$, yielding supertubes in $D=4$ and $D=6$ in addition to
the supertube in $D=10$. The 10-dimensional supertube is an effective
description of a configuration of IIA string theory that has an
alternative low-energy description as a supertube solution of IIA 
supergravity \cite{Emparan:2001ux}. Is there a similar `microscopic'
interpretation of the $D=4,6$ supertubes? 

As the starting point was a
membrane in $D=5,7$ we should first ask whether there are
supersymmetric theories in these dimensions that admit membrane
solutions. For $D=7$ the obvious candidate is a supersymmetric
Yang-Mills (SYM) theory because an instanton solution of the YM equations on
$R^4$ can be interpreted as a 1/2 supersymmetric membrane. For $D=5$
there are various candidates, one being a supersymmetric sigma-model 
because, for an appropriate choice of (necessarily hyper-K\"ahler) 
target space, there is a 1/2 supersymmetric lump
soliton that has a 5D interpretation as a membrane.

We should next ask whether these field theories admit 1/4
supersymmetric solutions that can be interpreted as intersections
of the 1/2 supersymmetric membranes. If so, the procedure
outlined above should yield solutions of the dimensionally-reduced
theories (6D SYM or 4D sigma models) that we could call 
{\it field theory supertubes}. These would have an effective 
description in terms of the supertube solution of the DBI action 
for a 6D or 4D membrane, just as the IIA supergravity supertube
has an effective description in terms of the DBI action for the
D2-brane (the Born-Infeld vector potential arising, in each case,
from dualization of a worldvolume scalar).

In the sigma-model case, the answer to this 
question is known. One can find an explicit 1/4 supersymmetric 
non-singular solution of a 5D sigma model that represents the intersection of 
two membranes (or 3-branes of the 6D sigma model) 
\cite{Portugues:2002ih}. A Scherk-Schwarz-type reduction to 
4D then yields a `massive' supersymmetric sigma model, and the intersecting
membrane solution of the massless 5D model becomes, 4D, the 1/4
supersymmetric `kink-lump', which can be interpreted as
a lump-string ending on a kink-membrane
\cite{Gauntlett:2000de}. A boost in the 5th dimension generalizes this to 
the `Q-kink-lump' of the massive 4D sigma-model \cite{Gauntlett:2000de}, 
and a boost to the speed of light yields a tubular  
configuration with a cross-section that is a 
1/4 supersymmetric Q-lump solution of the dimensionally-reduced 
3D massive sigma-model \cite{Leese:1991hr,Abraham:1992qv}. 
Thus, the Q-lump solution of 3D massive supersymmetric hyper-K\"ahler sigma
models is, when viewed as a tubular solution of the 4D sigma model,
a {\it field theory supertube}. In fact, it was this observation 
that led to the discovery of the
string theory supertube, and the above discussion is just a reversal
of the logic presented in \cite{Mateos:2001qs}. 

In this contribution I explore the same issues for SYM theories.  
Along the way, we will 
obtain a new one-parameter set of first-order equations for 
1/4 supersymmetric solutions 
of 6D SYM-Higgs theory. A limit of these equations, 
corresponding to a boost to the speed 
of light in the 7th dimension, yields equations that 
generalize the dyonic instanton equations \cite{Lambert:1999ua}  
of 5D SYM theory. Certain solutions of  these equations are interpreted 
as Yang-Mills supertubes, and a realization of them as IIB 
string theory configurations is 
suggested. I conclude with a discussion of some issues 
raised by these results.

\section{Yang-Mills Supertubes}

Let $F=dA + i[A,A]$ be the YM field-strength 2-form for YM 1-form 
potential $A$, which is a traceless
hermitian $2\times 2$ matrix for gauge group $SU(2)$. In the 
gauge $A_0=0$, any static bosonic 
solution of 7D SYM theory is solution of the
Euclidean YM equations on $\bR^6$. The generic solution of this 
type preserving 1/4 supersymmetry must satisfy a set of first order
differential equations, and one can choose coordinates
$x^1,x^2,\dots,x^6$ such that these first-order equations 
are \cite{Bak:2002aq}
\bea\label{BLPeqs}
F_{13} + F_{42} = 0 , \qquad && \qquad F_{14} + F_{23} = 0,  \nn
F_{15} + F_{62} = 0 , \qquad && \qquad F_{16} + F_{25} = 0,   \nn
F_{35} + F_{64} = 0 , \qquad && \qquad F_{36} + F_{45} = 0,  
\eea
\bea
F_{12} + F_{34} + F_{56} &=& 0. \nonumber
\eea
These equations are
equivalent to the Donaldson-Uhlenbeck-Yau  equations for Euclidean 
YM fields on $\bC^3$, and have been studied previously 
in the context of SYM theory (e.g., \cite{Figueroa-O'Farrill:1997iv}),
although not in the context of solitons of 7D SYM theory.

Of course, there will be special solutions of these equations that
preserve more than 1/4 supersymmetry. Apart from the vacuum, these are
the solutions for which $F$ is non-zero only 
on a 4-dimensional subspace of $\bR^6$; e.g., the `1234' subspace, 
in which case the equations reduce to 
\be
F_{13} + F_{42} = 0, \qquad 
F_{14} + F_{23} = 0, \qquad
F_{12} + F_{34} = 0,
\ee
which are equivalent to the self-duality equations
\be\label{SDeqs}
F_{ij} + {1\over2}\, \varepsilon_{ijkl} F_{kl} =0, 
\qquad (i,j,k,l=1,2,3,4). 
\ee
The solutions are instantons which, as mentioned above, can be
interpreted as 1/2 supersymmetric membrane solitons of the 7D 
SYM theory. If a 1/4 supersymmetric solution of the equations 
(\ref{BLPeqs}) is such that $F$ has support, 
asymptotically, on some 
4-plane then we would interpret this 4-plane as the space transverse 
to a membrane. Thus, it is reasonable to expect that, {\it for appropriate 
boundary conditions}, solutions of the equations (\ref{BLPeqs}) 
represent {\it intersecting} membranes. 

Let us now compactify one space dimension on a circle; take it to be
the $x^6$ direction, so that $A_6 = \Phi$, an adjoint
Higgs field. Take the YM fields to be independent of $x^6$; this means 
that $F_{56} = D_5\Phi$, where $D_5$ is the 5th component of the 
gauge-covariant derivative. Let $D_i$ ($i=1,2,3,4$) be the other 
four components. The equations (\ref{BLPeqs}) may now be written as
\bea\label{5deqs}
F_{ij} + {1\over2}\, \varepsilon_{ijkl} F_{kl} &=& -  \Omega_{ij}
D_5\Phi \nn
F_{i5} &=& \Omega_{ij} D_j \Phi,  
\eea
where $\Omega_{ij}$ are the entries of the $4\times4$ antisymmetric 
matrix with non-zero entries
\be
\Omega_{12} = -\Omega_{21} = \Omega_{34} = - \Omega_{43} = 1. 
\ee
We are still considering 
static solutions so it is understood that $A_0=0$, and that all 
fields are time-independent; in gauge-invariant terms,
\be
D_0\Phi =0, \qquad F_{05} =0, \qquad F_{0i}=0 \qquad (i=1,2,3,4). 
\ee
This means that the Gauss-law constraint
\be\label{glaw}
D_5 F_{05} + D_i F_{0i} =0
\ee
is trivially satisfied.

Let us suppose $A_3=A_4=0$, and that all fields become independent 
of  $x^3$ and $x^4$, 
asymptotically as $(x^3)^2 + (x^4)^2\rightarrow \infty$. 
In this case, equations (\ref{5deqs}) reduce to 
\be
F_{12} + D_5\Phi = F_{51} + D_2\Phi = F_{25} + D_1\Phi=0\, ,
\ee
which are the equations for a magnetic-monopole membrane in the $x^3,x^4$ 
plane (assuming that $\Phi$ is non-zero in the vacuum). If, on the other hand, 
$A_5=0$  asymptotically, as $(x^3)^2 + (x^4)^2 \rightarrow 0$, 
such that $\Phi$ becomes independent of $x^5$, and the fields 
$A_i$ ($i=1,2,3,4$) have an $x^5$ dependence such that 
$\partial_5 A_i = - \Omega_{ij}D_j\Phi$, then we 
are left with the self-duality equations (\ref{SDeqs}), and hence 
an instanton string in the $x^5$ direction. The instanton core of 
this string would collapse to a singularity if the string were 
not attached to the monopole membrane (which induces the
$x^5$-dependence of the YM fields). 
Thus, one expects there to exist a non-singular solution of the equations 
(\ref{5deqs}) with an  interpretation as an instanton-string 
ending on a  monopole domain wall\footnote{This interpretation 
was developed in unpublished work with Jerome Gauntlett and 
David Tong, following  the work in \cite{Portugues:2002ih} 
in which it was shown that 
a similar interpretation is indeed realized by the kink-lump solution of the
analogous sigma-model equations.}. 

We dimensionally reduced the 7D SYM theory in the $x^6$ direction,
allowing for a non-zero vacuum value for the Higgs field $A_6\equiv
\Phi$. Now, returning temporarily to the 7D perspective, we boost 
to velocity $v$ along $x^6$. This takes the equations (\ref{5deqs})
into the new set of equations
\bea\label{dyonic}
F_{ij} + {1\over2}\, \varepsilon_{ijkl} F_{kl} + \sqrt{1-v^2}\, 
\Omega_{ij}D_5\Phi &=& 0, \qquad 
F_{i5} - \sqrt{1-v^2}\, \Omega_{ij} D_j \Phi =0 \nn
F_{0i} + v D_i \Phi =0, \qquad  F_{05} + vD_5\Phi &=& 0 , \qquad 
D_0\Phi =  0. 
\eea
To determine the fraction of the 16 supersymmetries of the 6D SYM-Higgs
vacuum that are preserved by solutions of
these equations, it is convenient to note that the 7D SYM theory 
from which we started is the dimensional
reduction on $T^3$ of 10D SYM theory, so any
solution of the equations (\ref{dyonic}) is also a solution 
of 10D SYM theory with $A_6=\Phi$ but $A_7=A_8=A_9=0$, and
a field strength $F_{\mu\nu}$ ($\mu,\nu=0,1,2,\dots,9$) 
that is independent of $x^6,x^7,x^8,x^9$. The number of
supersymmetries preserved by any such solution is the number of 
linearly independent real, chiral, constant, 10D spinors $\epsilon$ 
such that $F_{\mu\nu}\Gamma^{\mu\nu}\epsilon=0$, where $\Gamma^\mu$
are the 10D Dirac matrices. Use of 
the equations (\ref{dyonic}) leads to the conclusion that the
independent constraints satisfied by $\epsilon$ are 
\be
\Gamma^{1234}\epsilon = -\epsilon, \qquad
\left(\Gamma^{1256} + v \Gamma^{1250}\right) 
\epsilon = - \left(\sqrt{1-v^2}\right)\epsilon\, .
\ee
These constraints imply preservation of 1/4 supersymmetry. Note,
that this does not, by itself, imply that the YM field equations are satisfied;
for that we must also impose the Gauss law 
condition (\ref{glaw}).

Given a solution of the unboosted 1/4 BPS equations (\ref{5deqs})
representing an instanton string ending on a monopole domain wall,
there should exist a corresponding solution of the boosted 1/4 BPS
equations and Gauss law constraint that represents an instanton 
string ending on a {\it dyon}  domain
wall. This is because the limit that led previously to the equations
for a 1/2 BPS monopole now leads to the equations for a 1/2 BPS dyon. 
Suppose that we have such a solution for any $v$ and that we take
$v=1$; i.e., we boost to the speed of light. In this case, we will have
a solution of the equations 
\bea\label{DI}
F_{ij} + {1\over2}\, \varepsilon_{ijkl} F_{kl} &=& 0 \nn
F_{0i} +  D_i \Phi &=& 0 \nn
D_0\Phi &=& 0
\eea
and
\be\label{DI2}
F_{i5} =0 , \qquad F_{05} + D_5\Phi =0,
\ee
which are obtained from (\ref{dyonic}) by setting $v=1$. The equations
(\ref{DI}) are 1/4 BPS equations for a {\it dyonic instanton} 
\cite{Lambert:1999ua}. Given an instanton solution of the self-dual 
YM equations, the other dyonic instanton equations are solved by
setting $A_0=\Phi$ for time-independent Higgs field $\Phi$, in which
case (\ref{DI2}) is equivalent to
\be\label{a5}
 \partial_5 A_i = D_i A_5 \,  , \qquad \dot A_5=0,
\ee
and the Gauss law constraint becomes
\be\label{gl}
\left(\sum_{i=1}^4 D_i^2 + D_5^2\right)\Phi = 0.
\ee
That is, $\Phi$ must solve the covariant 5D Laplace equation in a YM background
provided by a 4D instanton, with $x^5$-dependence given in terms of $A_5$ 
by (\ref{a5}). 

If we suppose that $A_5=0$ then $\partial_5A_i=0$.  Assuming that  $\Phi$ 
is also  independent of $x^5$, we get a string-like solution of 6D 
SYM-Higgs theory, with a dyonic instanton core
that carries `electric' charge $Q$ in addition to instanton
number $N$ (it would be interesting to investigate whether more
general solutions are possible, but that will not be done here).  
For the sigma-model supertube, the core is a Q-lump, and
a Q-lump has an interpretation as a charged closed loop of kink-string 
\cite{Abraham:1992qv}; the analogous interpretation of the dyonic
instanton would be as a charged closed loop of monopole-string, and 
this interpretation is also suggested by various other 
arguments \cite{Kruczenski:2002mn,Bak:2002ke}. 
However, the number of monopoles in a given solution of 
the 1/2 BPS equations of 4D SYM-Higgs theory is
determined by the number of zeros of the Higgs field, and the
positions of these zeros are the positions of the monopole. 
A single monopole, or dyon, has a single zero of the 
Higgs field, which will lift to a line of zeros in $5D$; a monopole,
or dyon, loop will thus be associated with a closed loop of zeros of
the Higgs field. In contrast, dyonic instantons corresponding to
instantons found by the 't Hooft ansatz 
have only {\it isolated} zeros of the Higgs 
field \cite{Eyras:2000dg}.

It therefore appeared, until recently, that the interpretation of 
dyonic instantons as  charged loops of monopole-string could not 
be correct. However, recent work of Kim and Lee \cite{Kim:2003gj}
has shown that the locus of zeros of the Higgs field for a {\it
generic} dyonic instanton with instanton number $N\ge2$ is a closed
curve, exactly as one would expect for a loop of monopole-string. 
Already for $N=2$, for which the general instanton solution can be
found from the Jackiw-Nohl-Rebbi ansatz, there is an additional
parameter as compared to the  't Hooft ansatz, and this yields a
one-parameter family of closed curves that degenerate to two points in
the 't Hooft ansatz limit. Lifting to $D=6$ we have a configuration of 
6D SYM-Higgs theory in which the Higgs zero lie on a tube. This is a 
{\it field theory supertube}; as I have argued here,  
it is related to an intersecting membrane solution of 7D SYM 
theory in the same way that the D2-brane supertube of IIA string 
theory is related to a configuration of intersecting M2-branes of
M-theory.

\section{String Theory Realizations}

We have found a 6D Yang-Mills supertube by a procedure 
that is analogous to one 
that can be used to find the 10D string theory  supertube. However, 
there is also a {\it direct} connection between the two that arises 
from the interpretation of SYM-Higgs theory, for gauge group $SU(2)$, 
as the effective field theory on a pair of parallel D-branes in type
II 
string theory.

Let us first consider the 5D $SU(2)$ SYM-Higgs  theory on a pair 
of parallel D4-branes. 
A line of zeros of the Higgs field, corresponding to a
monopole-string, 
would have a natural  interpretation as the endpoint of a planar 
D2-brane since T-duality in a direction parallel to 
the line yields a D1-string stretched between two D3-branes, 
which is the standard D-brane 
realization of a BPS magnetic monopole. A closed loop of zeros 
of the Higgs field  therefore 
represents the (common) boundary of a tubular D2-brane on the 
two D4-branes. The non-zero 
instanton number $N$ indicates $N$
dissolved D0-branes, but D0-brane charge is magnetic charge on a 
D2-brane. Moreover, the 
fraction of supersymmetry preserved by the total D-brane 
congfiguration is expected to be 1/8, 
which translates to 1/4 of the supersymmetry of the SYM-Higgs 
theory vacuum. Thus,  the 
generic  $SU(2)$ dyonic instanton has a string theory 
interpretation \cite{Kim:2003gj} as a supertube stretched 
between two D4-branes\footnote{In the one instanton case, 
the loop of  Higgs zeros degenerates to a point, and the 
interpretation is as a IIA string stretched between  two 
D4-branes carrying D0-brane charge \cite{Zamaklar:2000tc}.}.  
By T-duality we can convert this to a 1/8 supersymmetric IIB 
string configuration in which a D3-brane with $S^1\times R$ boundary 
having $N$ dissolved $D1$ strings, is stretched (along with 
some number of dissolved IIB strings) between two D5-branes. 
The common boundary on the D5-branes is the tubular locus of zeros  
of the Higgs field of 6D SYM-Higgs theory; in other words,  
{\it we have a IIB string theory 
description of the  1/4 supersymmetric field-theory supertube 
of 6D SYM-Higgs theory}.  

If the supertubular D3-brane suspended between the D5-branes 
collapses, `precipitating' out the dissolved IIB and D1 strings  
then we end up with a configuration represented by the array
\be
\ba{lccccccccc}
D5: & 1 & 2 & 3 & 4 & 5 & -  & - & - &  - \\
D1: & 1 & -  & -  & -  & -  & -  & - & - & -  \\
F1:  & -  & -  & -  & -  & -  & 6 &  -& - & -  
\ea
\ee
where the first row represents the two D5-branes, in which the 
`precipitated'  D1-branes are actually still dissolved, and `F1' 
indicates the `fundamental' IIB  strings. By adding angular 
momentum, one can reverse the collapse and blow up the IIB and 
D1 strings to the configuration described previously. Let us now 
increase the IIB coupling and pass to the S-dual  IIB string 
theory, and then T-dualize in the $x^6$ direction; this yields 
a IIA configuration represented by the array
\be
\ba{lccccccccc}
KK: & - & - & - & - & - & o  & x & x &  x\\
F1: & 1 & -  & -  & -  & -  & -  & - & - & -  \\
D0: & -  & -  & -  & -  & -  & - &  -& - & -  
\ea
\ee
where `KK'  (for Kaluza-Klein monopole) indicates a 4-dimensional 
ALE space, and the circle in the 6th position indicates that 
$\partial_6$ is the $U(1)$ Killing vector field;  since we 
started with two D5-branes, this Killing vector field should 
have two isolated singularities. However, we could view this
ALE space as a local description of a compact K3-manifold. 
Expansion of the IIA string and D0-branes by the addition of 
angular momentum then yields a supertube in what is effectively 
a 6D spacetime obtained by K3 compactification of the IIA theory. 
This should be related to the M2-brane of K3-compactified M-theory 
in the same way that the original supertube is related to the 
M2-brane in the 11-dimensional Minkowski vacuum. As M-theory on 
K3 is dual to the heterotic string theory on $T^3$, this membrane 
is dual to the membrane of $T^3$-compactified heterotic string 
theory, which is just the heterotic fivebrane wrapped on the $T^3$. 
We may conclude  from this there must exist a supertube  of 
$T^4$-compactified heterotic string theory dual to the D2-brane 
supertube of IIA string theory. The corresponding 
supergravity/SYM solution is presumably the heterotic dyonic 
instanton \cite{Eyras:2000dg}.

\section{Fuzzy Conjectures} 

We have seen that a D2-brane supertube can be suspended between 
two D4-branes, and 
that there is a T-dual of this configuration  for which the  
interpretation within an effective 6D
SYM-Higgs theory is as a field theory supertube. Can the 
latter be suspended, preserving 
supersymmetry, between other branes? Yes, because the 
two D5-branes can themselves 
be suspended between two NS5-branes. Consider the 
`precipated' configuration represented 
by the 1/16 BPS array 
\be
\ba{lccccccccc}
D5:   & 1 & 2 & 3 & 4 & 5 & - & - & - &  - \\
NS5: & - & 2 & 3 & 4 & 5 & 6 & - & - & - \\
D1:   & 1 & -  & -  & -  & -  & - & - & - & -  \\
F1:    & -  & -  & -  & -  & -  & 6 & - & - & -  
\ea
\ee
where the second row represents the additional pair of 
NS5-branes, and the low energy 
effective field theory is the (1+4)-dimensional  
theory on the D5/NS5 intersection. The 
IIB strings are parallel to the NS5-branes and the 
D1-strings are suspended between 
them (like the D5-branes in which they are dissolved). 
Adding angular momentum to 
blow up the IIB and D1 strings to a supertubular 
D3-brane would yield a configuration 
of the type sought, which would presumably correspond to  
some `Q-lump-type' supertube 
of the 5D effective field theory. However, the relationship of 
this construction to
the explicit M-theory description of the sigma-model Q-lump 
given in \cite{Gauntlett:2000de} is 
not obvious. This may be because there are other possible 
field theory realizations of 
the 5D supertube. 

As reviewed above, the work of Kim and Lee \cite{Kim:2003gj} has shown
that the generic dyonic instanton for instanton number $N$ can be
interpreted as a D2-brane supertube suspended between two D4-branes. 
The instanton number $N$ corresponds to the number of dissolved 
D0-branes; this would be infinite for an infinite supertube but can be 
finite for one suspended between D4-branes. For $N=2$ there is a
dyonic instanton solution for which the Higgs field zeros lie on a
circle. This corresponds to a supertube with circular cross-section. 
The limiting procedure explained here for finding a D2-brane 
supertube from the 11-dimensional supermembrane yields a supertube of
this type (because until the limit is taken 1/4 supersymmetry implies
a circular cross-section). However, 1/4
supersymmetry allows supertubes with other cross-sectional 
shapes \cite{Bak:2001xx}; in 
fact, any curve, even a non-planar one, is permitted \cite{Mateos:2001pi}. 
In contrast, the locus of Higgs
field zeros for the generic dyonic instanton with instanton number $N$
involves only a finite number of parameters, so it cannot correspond
to the generic classical supertube cut-off by D4-branes. On 
the other hand, it seems
likely that any cross-sectional curve will be possible in the limit as
$N\rightarrow\infty$, so that this should be viewed as a
semi-classical limit. In effect, the Yang-Mills supertube for finite
$N$ is a fuzzy supertube, analogous to the M(atrix) model supertube of 
\cite{Bak:2001kq} but constructed from a finite number of D0-branes.
This expectation suggests  a purely mathematical conjecture about 
the the locus of zeros of an adjoint Higgs field satisfying a covariant 
Laplace equation on $\bR^4$ in a YM instanton background with 
instanton number $N$.
Specifically, although imprecisely, it suggests that {\it this 
locus can be chosen to approximate any closed curve in $\bR^4$ 
with an error that goes to zero as $N\rightarrow \infty$}.  

Finally, it should be noted that there are Donaldson-Uhlenbeck-Yau equations for 
Euclidean YM theory on $\bE^8$ which are equivalent to static soliton equations 
for D=9 SYM theory that preserve 1/8 of the supersymmetry of the SYM vacuum, and 
there are other first order equations (comprehensively analysed 
in \cite{Bak:2002aq}) 
that imply preservation of 1/16 supersymmetry. It is possible that there exist 
non-singular solutions of these equations that could be interpreted  as triple 
or quadruple intersections of orthogonal 4-branes with instanton
cores, in analogy 
with the sigma-model case \cite{Portugues:2002ih}. 
These intersecting-instanton solutions 
will be unstable against collapse to a singularity in the 
Higgs phase, but in that case 
we have 1/2 supersymmetric monopole 5-branes, on which an (otherwise singular)
instanton 4-brane could have a boundary (this being the 
lift to 9D of the 5D instanton 
string ending on the 5D monopole membrane). This configuration of D=9 SYM-Higgs 
theory would preserve 1/4 supersymmetry but one can envisage many more 
complicated intersecting soliton configurations that preserve 
only 1/8 or 1/16 supersymmetry. 
There is still a lot to learn about supersymmetric intersections 
of field theory solitons. 
Field theory supertubes constitute just one element in a much 
larger, and still emerging, picture.

\vskip 1cm
{\bf Acknowledgements}:
I thank Jerome Gauntlett and David Tong for collaboration long ago on
some of the topics discussed here, and Olaf Lechtenfeld for bringing
to my attention the connection to the Donaldson-Uhlenbeck-Yau
equations.

\end{document}